\documentclass[3p,times,preprint]{elsarticle}

\usepackage{amssymb}
\usepackage{graphicx}
\usepackage{natbib}
\usepackage{epstopdf}
\usepackage{graphicx}
\usepackage{caption}
\usepackage{verbatim}
\usepackage{subcaption}
\usepackage{bbm}
\usepackage{amsmath}
\usepackage{subcaption}
\usepackage{caption}

\usepackage[figuresright]{rotating}

\begin{document}

\begin{frontmatter}


 \title{Title\tnoteref{label1}}





\author[label1,label4]{B. S. Collyer\corref{mycorrespondingauthor} }
\ead{benjamin.collyer@gmail.com}
\address[label1]{Centre for Complexity Science, University of Warwick, Coventry CV4 7AL, UK}
\author[label1,label2,label4]{C. Connaughton}
\address[label2]{ Mathematics Institute, University of Warwick, Coventry CV4 7AL, UK}
\author[label3]{D. A. Lockerby }
\address[label3]{ School of Engineering, University of Warwick, Coventry, CV4 7AL, UK}
\address[label4]{London Mathematical Laboratory, 14 Buckingham Street, London WC2N 6DF, UK}

\cortext[mycorrespondingauthor]{Corresponding author}


\title{Importance Sampling Variance Reduction for the Fokker-Planck Rarefied Gas Particle Method}




\begin{abstract}
Models and methods that are able to accurately and efficiently predict the flows of low-speed rarefied gases are in high demand, due to the increasing ability to manufacture devices at micro and nano scales. One such model and method is a Fokker-Planck approximation to the Boltzmann equation, which can be solved numerically by a stochastic particle method. The stochastic nature of this method leads to noisy estimates of the thermodynamic quantities one wishes to sample when the signal is small in comparison to the thermal velocity of the gas. Recently, Gorji et al have proposed a method which is able to greatly reduce the variance of the estimators, by creating a correlated stochastic process which acts as a control variate for the noisy estimates. However, there are potential difficulties involved when the geometry of the problem is complex, as the method requires the density to be solved for independently.

Importance sampling is a variance reduction technique that has already been shown to successfully reduce the noise in direct simulation Monte Carlo calculations. In this paper we propose an importance sampling method for the Fokker-Planck stochastic particle scheme. The method requires minimal change to the original algorithm, and dramatically reduces the variance of the estimates.
We test the importance sampling scheme on a homogeneous relaxation, planar Couette flow and a lid-driven-cavity flow, and find that our method is able to greatly reduce the noise of estimated quantities. Significantly, we find that as the characteristic speed of the flow decreases, the variance of the noisy estimators becomes independent of the characteristic speed.
\end{abstract}



\end{frontmatter}


\section{Introduction}
Recent technological advances have resulted in manufacturing processes that have made possible the production of mechanical devices that operate on the scale of micro and nanometers \cite{Reese2003}. Such technologies include lab-on-a-chip devices, micro-heat exchangers, gas chromatographers and micro-jet actuators for control in aerospace. At such small scales, the Navier-Stokes-Fourier (NSF) equations are no longer able to accurately model gas flows, due to the length scales of macroscopic gradients approaching the length of the molecules mean free path, $\lambda$. This results in the existence of a region known as the Knudsen layer near solid wall boundaries where the gas is prevented from relaxing to thermodynamic-equilibrium, invalidating the assumption that locally the gas is close to thermal equilibrium required for the NSF equations to be valid.

The Boltzmann equation is a mesoscopic model that is considered to provide the most accurate description of rarefied gases beyond Newton's laws. Before the advent of such small scale technologies, rarefied gas flows' largest application area was that of supersonic atmospheric flows, where the Mach number of flow, $\textrm{Ma}>1$. Currently, the prevalent method for numerically approximating the solution to the Boltzmann equation in such regimes is a stochastic particle method called direct simulation Monte Carlo (DSMC) \cite{Bird1994} \cite{Oran1998}. Due to the stochastic nature of the method, DSMC becomes very inefficient for low-speed flows. Typically the Mach number, $\textrm{Ma}\ll 1$ for flows within micro and nano technologies, and for a given level of statistical error, the computational costs of DSMC scale as $\textrm{Ma}^{-2}$ \cite{Hadjiconstantinou2003}. This results in very long computation times for such calculations, and methods which are able to efficiently solve for low speed flows are highly desirable.

Currently, there are two methods that are able to greatly reduce the variance of the desired thermodynamic outputs of DSMC calculations. The first, low-variance DMSC (LVDSMC), works by adapting the DSMC collision routine to calculate the evolution of the deviation $f_d=f - f_{M}$ from a Maxwellian distribution $f_{M}$ \cite{Homolle2007}. In low speed flows the deviation from equilibrium is small, allowing for a dramatic decrease in the variance of samples.  An alternative method, variance reduced DSMC (VRDSMC), is able to work without significant changes to the DSMC algorithm \cite{Al-Mohssen2010}. The method relies on importance sampling, which allows the algorithm to sample from an equilibrium distribution where the thermodynamic variables are known aproiri, to create estimators with smaller variance.

More recently an alternative method to DSMC, where the Boltzmann collision operator is approximated by a Fokker-Planck operator, has been developed and shown to be more efficient than the basic DSMC algorithm \cite{Jenny2010} \cite{Gorji2011}. Like DSMC, it is solved stochastically using notional particles that represent a certain number of real particles in the gas to be simulated, and as such, the basic algorithm suffers from the same inhibitive scaling with the Mach number. Recently Gorji et al. \cite{Gorji2015} have proposed a method to reduce the variance of the Fokker-Planck solution algorithm that relies on creating a correlated equilibrium solution using the same set of random numbers that are used in the stochastic solution of the non-equilibrium process. The parallel correlated equilibrium process is in effect a control variate for the non-equilibrium process.

This purpose of this paper is to develop an importance sampling variance reduction scheme for the Fokker-Planck method and demonstrate its effectiveness in simple test cases. The paper is organised in the following way: in section 2 we introduce the Fokker-Planck model, and numerical stochastic particle scheme which we would like to create variance reduced estimators for.  We then outline the general method that allows one to create variance reduced estimators by exploiting known information about how the macroscopic fields behave at equilibrium. In section 3 we describe the variance reduction scheme proposed by Gorji el al. \cite{Gorji2015}, which creates a correlated equilibrium scheme. In section 4 we propose our importance sampling scheme, which we test in section 5 on a homogeneous relaxation, Couette flow and a lid-driven cavity flow. We then compare the importance sampling method against the results obtained by using a correlated equilibrium solution.

\section{Background}
\subsection{The Fokker-Planck collision operator}
The Fokker Planck collision operator has appeared in several different contexts, originally derived for the distribution function of a Brownian particle in a fluid \cite{Chandrasekhar1943}, but can also be derived from an expansion of a linear Boltzmann equation, when considering the evolution of density function for a particle in a heat bath \cite{Ferrari1982232}. It has been used to model electrons, dense liquids and more recently has received attention for its ability to model rarefied gas flows \cite{Cercignani1988}. More recently it has been extended to describe flows of monatomic gas mixtures \cite{Jenny2012}, diatomic molecules \cite{Gorji2013} and has been coupled to DSMC \cite{Gorji2014}. The equation for the one-particle distribution function $f(\mathbf{x},\mathbf{v},t)$, over a state-space comprised of the position $\mathbf{x}\in\mathbb{R}^3$, velocity  $\mathbf{v}\in\mathbb{R}^3$ and time $t\in\mathbb{R}^+$ takes the form:

\begin{align}
\frac{\partial f}{\partial t} + \mathbf{v}\cdot \nabla_\mathbf{x}f&= \mathcal{A}(f)\\
&:=  \frac{1}{\tau} \nabla_\mathbf{v}\cdot\Big[\mathbf{c}f + RT \nabla_\mathbf{v}f \Big] \label{eq:fp}, 
\end{align}
where $\tau$ is a relaxation time, $\mathbf{c} = \mathbf{v} - \mathbf{u}$ is the local relative molecular velocity, $\mathbf{u}$ is the mean velocity:
\begin{align}
 \mathbf{u}(\mathbf{x},t) = \frac{1}{\rho} \int \mathbf{v}f(\mathbf{x},\mathbf{v},t)\,\textrm{d}\mathbf{v},
\end{align}
$T$ is the local temperature given by
\begin{align}
T(\mathbf{x},t) =\frac{1}{3R\rho} \int c^2f(\mathbf{x},\mathbf{v},t)\,\textrm{d}\mathbf{v} ,
\end{align}
where $R$ is the specific gas constant and $\rho$ is the local density given by
\begin{align}
\rho(\mathbf{x},t) =\int f(\mathbf{x},\mathbf{v},t)\,\textrm{d}\mathbf{v}.
\end{align}
The collision operator $\mathcal{A}$ has the property of conserving mass, momentum and energy. That is 

\begin{align}
\int \mathcal{A}(f)\mathbf{\psi}\,d\mathbf{v} = 0,
\end{align}
where $\psi = \{1,\mathbf{v},v^2\}$ is the set of collisional invariants. The advantage of having a collision operator which can be written as a Fokker-Planck equation is that there exists an equivalent stochastic differential equation (SDE) representation for the dynamics of a random variable $\{\mathbf{X}_t,\mathbf{V}_t\}$ whose distribution $f$ evolves according to (\ref{eq:fp}):

\begin{align}
\textrm{d}\mathbf{X}_t &= \mathbf{V}_t \textrm{d}t\\
\textrm{d}\mathbf{V}_t &= \frac{1}{\tau}(\mathbf{V}_t-\mathbf{U})\textrm{d}t + \sqrt{\frac{2RT}{\tau}}\,\textrm{d}\mathbf{W}_t \label{eq:SDEs},
\end{align}
where $\mathbf{W}_t$ is a 3-dimensional Wiener process. An efficient scheme for evolving a collection of representative particles with positions and velocities $\{\mathbf{X}^j(t),\mathbf{V}^j(t)\}$, $j=1\ldots N$, distributed according to the distribution $f(\mathbf{x},\mathbf{v},t)$ in time was devised by \emph{Jenny et al.} \cite{Jenny2010}, and can be summarised as:

\begin{align}
V_i\left(t+\Delta t\right) &= V_i(t) - \left(1-e^{\Delta t/\tau}\right)\Big(V_i(t) - U_i(t)\Big)+\sqrt{\frac{C^2}{B}\xi_{1,i}}+ \sqrt{A-\frac{C^2}{B}}\xi_{2,i}\label{eqn:update}\\
X_i(t+\Delta t) &= X_i(t) + U_i(t)\Delta t + \tau\Big(V_i(t)-U_i(t)\Big)\left(1-e^{-\Delta t/\tau}\right)+\sqrt{B}\xi_{1,i},
\end{align}
where $i=1,2,3$ indexes the dimension,

\begin{align}
A &= RT\left(1-e^{-2\Delta t/\tau}\right)\\
B &= RT\tau^2\left(\frac{2\Delta t}{\tau} - \left(1-e^{-\Delta t/\tau}\right)\left(3-e^{-\Delta t/\tau}\right)\right)\\
C&=RT\tau\left(1-e^{\Delta t/\tau}\right)^2,
\end{align}
and $\Delta t$ is the time-step, $\tau$ is the relaxation time and $\xi$ are standard normal distributed random variables. The spatial domain is gridded into cells, and expectations of macroscopic quantities of interest are calculated during each time-step for each computational cell. The correct viscosity is obtained by choosing the relaxation time $\tau = 2\mu/p$, where $\mu$ is the viscosity and $p$ is the pressure.

\subsection{Variance reduction for Monte Carlo sampling}\label{sec:types_paper}

In this section we outline the general framework which allows one to reduce the variance of an estimator of a particular random variable, common to control variate and importance sampling schemes. Essentially, both methods exploit information about errors in estimates of known quantities. The general principal is as follows. Suppose we have a random variable $X$, and we wish to estimate $\mathbb{E}[X]$, let our estimate of $\mathbb{E}[X]$ be donated by $\hat{X}$. Let $Y$ be a different random variable with known expectation $\mathbb{E}[Y]$ with an estimator denoted by $\hat{Y}$. Then for any $\alpha\in\mathbb{R}$ we can use the identity

\begin{align}
\mathbb{E}[X] = \mathbb{E}[X + \alpha Y] - \alpha \mathbb{E}[Y],
\end{align}
to create a new unbiased estimator for $\mathbb{E}[X]$,

\begin{align}
{X}_{VR} = \hat{X} + \alpha \hat{Y} - \alpha \mathbb{E}[Y]. \label{eq:vr}
\end{align}
The variance of this estimator is

\begin{align}
\textrm{Var}[{\hat{X}_{VR}}] = \textrm{Var}[\hat{X}] +\alpha^2\textrm{Var}[ \hat{Y}] + 2\alpha\,\textrm{Cov}[\hat{X},\hat{Y}]
\end{align}
and if we minimise this over possible choices of $\alpha$, then minimiser $\alpha^*$is given by 

\begin{align}
\alpha^*= -\dfrac{\textrm{Cov}[\hat{X},\hat{Y}]}{\textrm{Var}[\hat{Y}]},
\end{align}

hence the variance for this choice of $\alpha$ is 

\begin{align}
\textrm{Var}[{\hat{X}_{VR}}] = \textrm{Var}[\hat{X}] - \dfrac{\textrm{Cov}[\hat{X},\hat{Y}]^2}{\textrm{Var}[\hat{Y}]}.
\end{align}

The only condition required for the variance of the estimator to be less than the variance of the original estimator is for $\textrm{Cov}[\hat{X},\hat{Y}]>0$, and so $\hat{X}$ and $\hat{Y}$ being dependent is a necessary condition. This is all supposing that we already know, $\alpha^*$ which presupposes that we already know $\textrm{Cov}[\hat{X},\hat{Y}]$. In general this is not something that is not known a priori, but can be estimated throughout the simulation. In the next sections we will see how in practice it is possible to exploit this.

\section{Parallel process variance reduction}
We now briefly describe the method proposed by \cite{Gorji2015}. The objective of the method is to create a stochastic process $\mathbf{Z}_t$ that is able to run in parallel to the original particle scheme, where crucially, the macroscopic fields are already known. If this is performed in a manner where the parallel process is correlated with the original stochastic process then the variance of the estimators can be reduced in the way described in the previous chapter. The coupling of the new stochastic process $\mathbf{Z}_t$ to the original SDEs (\ref{eq:SDEs}) is achieved in the following way:

\begin{align}
\textrm{d}\mathbf{X}_t &= \mathbf{V}_t \textrm{d}t\\
\textrm{d}\mathbf{V}_t &= \frac{1}{\tau}(\mathbf{V}_t-\mathbf{u})\textrm{d}t + \sqrt{\frac{2RT}{\tau}}\,\textrm{d}\mathbf{W}_t\\
\textrm{d}\mathbf{Z}_t &= \mathbf{A} \textrm{d}t + D\,\textrm{d}\mathbf{W}_t
\end{align}
where the coefficients $\mathbf{A}$ and $D$ are chosen to keep the marginal distribution of $(\mathbf{X}_t,\mathbf{Z}_t)$, which we denote $f_0(\mathbf{x},\mathbf{z},t)$, as a solution to a Fokker-Planck equation

\begin{align}
\dfrac{\partial f_0}{\partial t} +  \nabla_\mathbf{x}\cdot(\mathbf{U}f_0) =  \nabla_\mathbf{v}\cdot\Big[\mathbf{A}f_0 + \frac{D^2}{2} \nabla_\mathbf{v}f_0\Big] \label{eq:fp2}, 
\end{align}
which when supplemented by appropriate boundary conditions, is solved by a Maxwellian

\begin{align}
f_M(\mathbf{x},\mathbf{z},t) = \frac{\rho(\mathbf{x},t)}{\left(2\pi RT_0\right)^{3/2}}\exp\left(-\frac{z^2}{2RT_0}\right). \label{mb}
\end{align}
The way to choose $\mathbf{A}$ and $D$ to ensure that (\ref{mb}) is a solution of (\ref{eq:fp2}) is discussed in depth in \cite{Gorji2015}. The coupling of the processes $\mathbf{Z}_t$ and $\mathbf{V}_t$, by the same Wiener process $\mathbf{W}_t$ requires that when we use particles to generate numerical solutions, the same set of random numbers is used for both the equilibrium and non-equilibrium processes. This results in the correlation of estimations of expected quantities, allowing the kind of variance reduction outlined in the previous section. In this method, the parallel equilibrium process, with distribution $f_M$, shares the same density $\rho$ as the non-equilibrium process described which has $f$ as its distribution function, and so the method cannot reduce the noise in the density calculation directly. Gorji et al propose that the density $\rho$ is found from the continuity equation using conventional finite difference methods. The results they obtain for a homogeneous relaxation, Poisuille flow and lid-driven cavity flows show the method has the ability to substantially reduce noise. Because this method uses common random numbers to reduce the variance, we will refer to this method as a common random numbers (CRN) scheme.

\section{Importance sampling}
The method we propose is an importance sampling scheme. It differs from the importance sampling scheme utilised by VRDMSC \cite{Al-Mohssen2010}, because the DSMC and Fokker-Planck method account for collisions in different ways. The principle that underpins the importance sampling remains the same however. Suppose we are interested in evaluating the expectation of $g(\mathbf{v})$ where v is distributed according to the distribution function $f$, then given $N$ independent samples $\left\{ \mathbf{V}_1,\ldots, \mathbf{V}_N \right\}$ distributed according to $f$ the following definition gives rise to the estimate:

\begin{align}
\mathbb{E}_f[g(\mathbf{v})] &:= \int g(\mathbf{v})f(\mathbf{v})\,d\mathbf{v}\\
& \approx \dfrac{1}{N}\sum_{i=1}^{N}g(\mathbf{V}^i),
\end{align}
which we know from the Central Limit Theorem, has an error of order $N^{-1/2}$. We now define a weight function

\begin{align}
W(\mathbf{v}) := \dfrac{f_{\textrm{ref}}(\mathbf{v})}{f(\mathbf{v})},
\end{align}
which is a measure of how likely one is to see a particle with velocity $\mathbf{v}$, relative to how likely one is to observe this particle if it was distributed to a reference density $f_{\textrm{ref}}$. This definition is well defined if the distribution $f$ is absolutely continuous with respect to $f_{\textrm{ref}}$, meaning that $f_{\textrm{ref}}(S)=0$ whenever $f(S)=0$ for any subset $S$ of the state-space. This definition can be viewed as a Radon Nikodym derivative. It can then be observed that the expectation of $g(\mathbf{v})$ with respect to the reference distribution can be estimated using the original samples:

\begin{align}
\mathbb{E}_{f_{\textrm{ref}}}[g(\mathbf{v})] &= \int f_{\textrm{ref}}(\mathbf{v})g(\mathbf{v})\,d\mathbf{v}\\
& = \int f(\mathbf{v})\dfrac{f_{ref}(\mathbf{v})}{f(\mathbf{v})}g(\mathbf{v})\,d\mathbf{v}\\
& = \int f(\mathbf{v})W(\mathbf{v})g(\mathbf{v})\,dv\\
& \approx \dfrac{1}{N}\sum_{i=1}^{N}W(\mathbf{V}^i)g(\mathbf{V}^i).
\end{align}
This is significant as it allows one to sample from the reference distribution $f_{\textrm{ref}}$, using the original set of samples from the distribution $f$. If the reference distribution is Maxwellian, $f_{\textrm{ref}}=f_{\textrm{M}}$ where one knows the thermodynamic fields analytically, then one has the ability to create variance reduced estimators as described in section 2. In order to practically apply this, we need a method of evolving the weights and velocities $\{\mathbf{\mathbf{V}}^i,W^i\}$ in time, where $W^i = W(\mathbf{\mathbf{V}}^i)$. For VRDSMC this is possible because it can be shown directly from the Boltzmann equation, that if two particles are chosen to collide with weights $W_i$ and $W_j$ then the post collision weights must be equal to $\frac{1}{2}(W^i+W^j)$. Because the Fokker-Planck dynamics have no explicit collisions, a different way to update the weights is needed.

Importance weights can be initialised exactly, because the initial velocities of the particles are distributed according to a prescribed initial distribution $f_0$. As time is evolved during the calculation, the distribution of velocities will change and hence so must the weights attached to each particle. VRDSMC is able to do this by creating collision rules that ensure that post collision velocities are still able to sample from the same reference distribution. However, these rules are not relevant for the Fokker-Planck particle dynamics as there are no explicit collisions.

\subsection{Weight update rule}

Instead, let us suppose that a given particle updates its velocity from $\mathbf{V}_t\rightarrow\mathbf{V}_{t+1}$, where $\mathbf{V}_t$ is distributed according to $f_t$ and $\mathbf{V}_{t+1}$ is distributed according to $f_{t+1}$, and that we know $W_t = W(\mathbf{V}_t)$. In order to update the weight exactly, one would need to know $f_{t+1}(\mathbf{V}_{t+1})$, however this distribution function is unknown. A simple method to estimate the updated weight is to use the zeroth order Taylor expansions of the joint distributions of $\mathbf{V}_t$ and $\mathbf{V}_{t+1}$, resulting in the estimate:

\begin{align}
W_{t+1}\approx\widehat{W}_{t+1}&:= \dfrac{ f_{eq}(\mathbf{\mathbf{V}_{t+1}}| \mathbf{V}_t)f_{eq}(\mathbf{V}_t)}{f_{t+1}(\mathbf{\mathbf{V}_{t+1}}| \mathbf{V}_t)f_t(\mathbf{V}_t)}\\
&= \dfrac{ f_{eq}(\mathbf{\mathbf{V}_{t+1}}| \mathbf{V}_t)}{f_{t+1}(\mathbf{\mathbf{V}_{t+1}}| \mathbf{V}_t)}W_t(\mathbf{V}_t). \label{eq:rule}
\end{align}
This has approximation immediately has some desirable properties. Firstly, the error of the approximation decays with $\Delta t$. Also, it is possible to calculate this explicitly from the update rule $\mathbf{V}_t\rightarrow\mathbf{V}_{t+1}$ given by equation (\ref{eqn:update}). This conditional distribution will be a gaussian centred on $\mathbf{V}_t$ plus the deterministic drift, with a temperature dependent variance. Further to this, it has the correct conditional expectation $\mathbb{E}[\widehat{W}_{t+1}|W_{t}]=W_{t}$ when the distribution is stationary. However, on its own it is not a suitable choice as if such a rule is repeated the variance of this approximation diverges, which is a common problem for this type of particle weight importance sampling method \citep{Al-Mohssen2010} \citep{Chun2005}. This is a problem, because to reduce the variance of our estimators in a meaningful way, we require the weights to be close to unity. To avoid this problem we use the same kernel density estimator approach as used by the VRDSMC method \citep{Al-Mohssen2010}. Kernel density estimation (KDE) is a method that allows one to obtain an estimate $\hat{f}$ of a density function $f$ from samples distributed according to that density function in the following way:

\begin{align}
\widehat{f}(\mathbf{v}) = \frac{1}{N}\sum_{i=1}^{N}K_r(\mathbf{v}-\mathbf{v}^i),
\end{align}
where $K_r$ is a kernel function that integrates over the state-space to 1, and $r$ is a smoothing parameter that controls the width of the kernel function. We use the same spherical kernels as \cite{Al-Mohssen2010}:

\begin{align}
K_r(\mathbf{v}-\mathbf{v}^i) = \left\{ 
  \begin{array}{l l}
    3/(4\pi r^3) & \quad \text{if $\|\mathbf{v}-\mathbf{v}^i\|<r$ }\\
    0 & \quad	 \text{otherwise}
  \end{array} \right. ,
\end{align}
which returns the reciprocal of the volume of a sphere of radius $r$ if $\mathbf{v}^i$ lies within the sphere of radius $r$ centred on $\mathbf{v}$, and otherwise returns a zero. If we combine this with (\ref{eq:rule}), the update rule that is obtained is

\begin{align}
W_{t+1}(\mathbf{V}^i) &\approx \dfrac{\sum_{j=1}^NK_r(\mathbf{V_i}-\mathbf{V}^j)\widehat{W}_{t+1}(\mathbf{V}^j)}{\sum_{j=1}^NK_r(\mathbf{V_i}-\mathbf{V}^j)}\\
&= \frac{1}{\big| S_r(\mathbf{V}^i)\big|}\sum_{\mathbf{V}^j\in S_r(\mathbf{V}^i)} \widehat{W}_{t+1}(\mathbf{V}^j),
\end{align}
where $S_r(\mathbf{V}^i)=\big\{ \mathbf{V}^j : \, \|\mathbf{V}^j-\mathbf{V}^i\|<r\big\}$ is the set of samples whose members lie within the sphere of radius of $r$ centred on $\mathbf{V}^i$. This KDE step has the effect of smoothing out the variation introduced by using a conditional probabilities to estimate a marginal probability, and making the scheme more stable. Increasing the smoothing parameter $r$ results in an estimator with a smaller variance, however it also increases the bias of the estimation, so ideally $r$ should be chosen to be as small as possible whilst maintaining an acceptable level of variation.

\subsection{Boundary Conditions}

We use the same boundary condition methodology as prescribed by the VRDSMC method, that is for diffusely reflecting fully accommodating walls, with temperature $T_{wall}$ and tangential velocity $u_{wall}$. Supposing that the Maxwellian distribution at the boundary is given by $f_{wall}(\mathbf{v}) = \rho_{wall}P_{MB}(\mathbf{v})$, where $P_{MB}$ is a gaussian probability density, and the boundary is the plane $x=0$, then the no flux boundary condition is given by

\begin{align}
\rho_{wall} \int\limits_{v_x>0}v_x P_{MB}(\mathbf{v})\,d\mathbf{v} +  \int\limits_{v_x<0}v_x f(\mathbf{v})\,d\mathbf{v}=0,
\end{align}
and similarly for the equilibrium solution 

\begin{align}
\rho_{wall,eq} \int\limits_{v_x>0}v_x P_{MB,eq}(\mathbf{v})\,d\mathbf{v} +  \int\limits_{v_x<0}v_x W(\mathbf{v})f(\mathbf{v})\,d\mathbf{v}=0.
\end{align}
The second integrals in the above equations are the particle fluxes, and can be estimated by counting the number of particles $N_{in}$ the cross through a wall of area $\Delta s$ in a time period $\Delta t$ by $(1/\Delta s\Delta t)N_{in}$, and at equilibrium is estimated by $(1/\Delta s\Delta t)\sum^{N_{in}}_i W_i $. Also, we can use the analytical properties of the Gaussian distribution to evaluate the first integrals:

\begin{align}
\int\limits_{v_x<0}v_x P_{MB}(\mathbf{v})\,d\mathbf{v}  = \frac{1}{\sqrt{2\pi}}\sqrt{\dfrac{kT}{m}}.
\end{align}
Therefore, a particle that changes velocity from $\mathbf{V}$ to $\mathbf{V}'$ when colliding with a wall, changes its weight according to

\begin{align}
W' = W(\mathbf{V}')& =\dfrac { f_{eq}(\mathbf{V}')}{f(\mathbf{V}')}\\
&=\dfrac{\rho_{wall,eq}P_{MB,eq}(\mathbf{V}')}{\rho_{wall}P_{MB}(\mathbf{V}')}\\
& = \sqrt{\dfrac{T_{wall}}{T_{wall,eq}}} \dfrac{\sum^{N_{in}}_i W_i}{N_{in}} \dfrac{P_{MB,eq}(\mathbf{V}')}{P_{MB}(\mathbf{V}')},
\end{align}
where typically, we choose the temperature of the equilibrium wall boundary condition to be equal to the temperature of non-equilibrium wall boundary boundary condition, i.e.  $T_{wall}=T_{wall,eq}$.

\section{Results}

\subsection{Homogeneous Relaxation to Equilibrium}

\begin{figure}
        \centering
        \begin{subfigure}[b]{0.45\textwidth}
                \includegraphics[width=\textwidth]{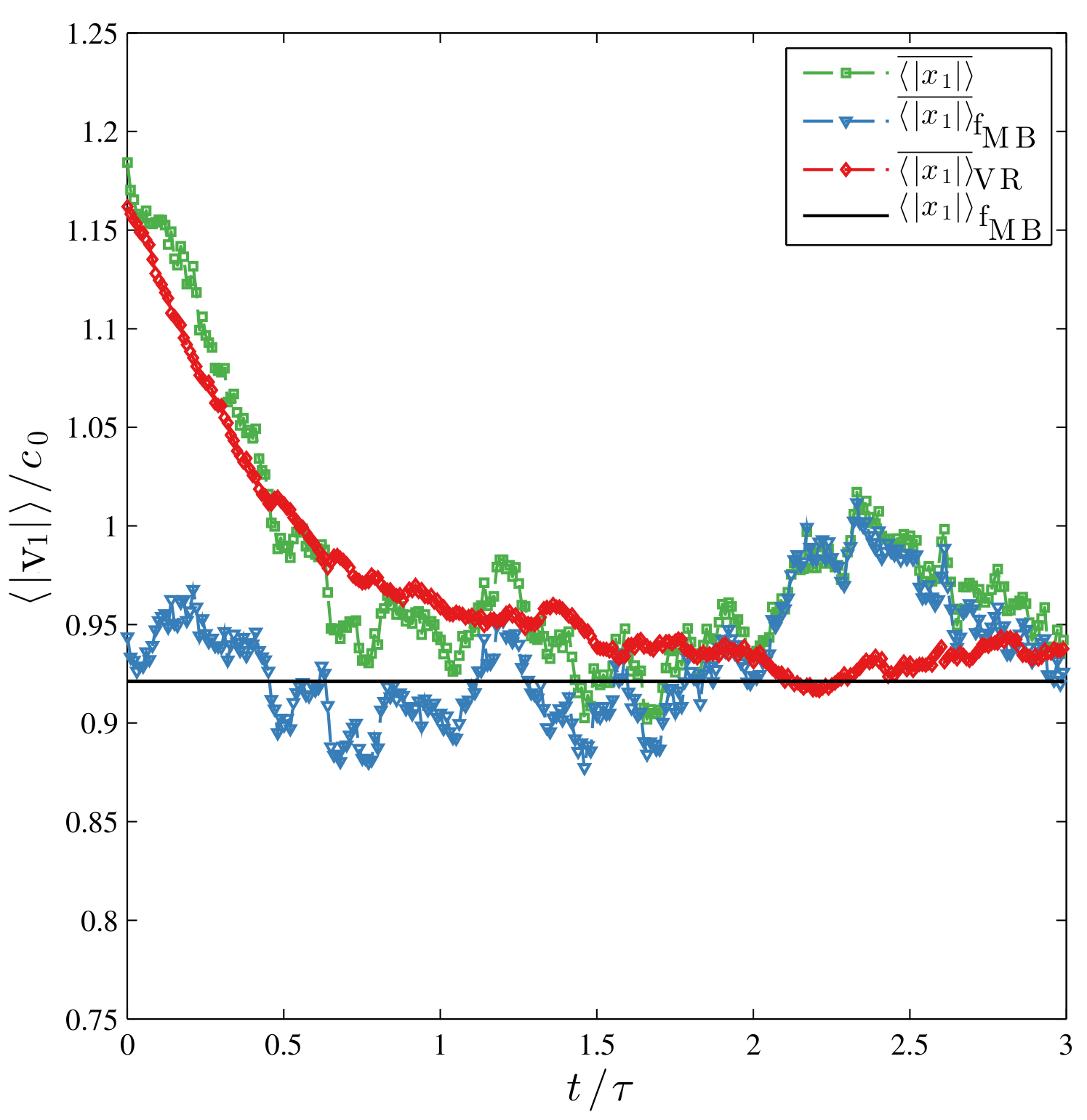}
                \caption{Without stabilisation}
                \label{fig:gull}
        \end{subfigure}%
        ~ 
        \begin{subfigure}[b]{0.45\textwidth}
                \includegraphics[width=\textwidth]{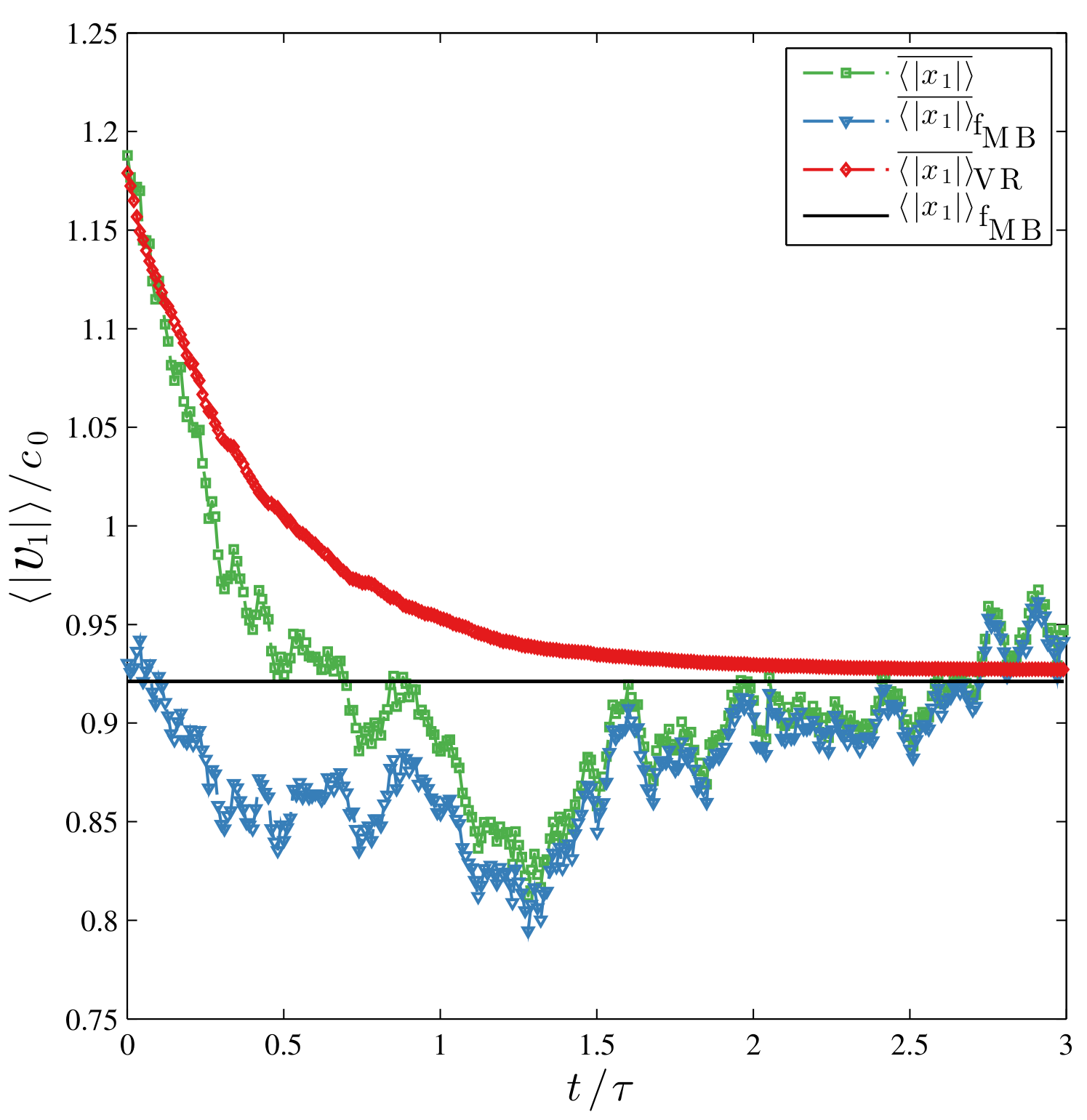}
                \caption{With stabilisation}
                \label{fig:tiger}
        \end{subfigure}
        \caption{Homogeneous relaxation towards equilibrium, (a) without KDE, (b) with KDE, smoothing parameter $r=0.05c_0$. The green squares represent the time-series of the standard MC estimator of the non-equilibrium process; t	he blue triangles represent the time-series of the standard MC estimator of the process biased to sample from equilibrium; the red diamonds represent the time-series of the VRFP estimator of non-equilibrium process; the black line represents the exact expectation at equilibrium.}
 \end{figure}

We will demonstrate the effectiveness of this method first with a homogeneous relaxation to equilibrium, i.e. when $f(t,\mathbf{x},\mathbf{v})=f(t,\mathbf{v})$ has no spatial component. We start from an initial distribution of particles 

\begin{align}
f_0(\mathbf{v})=(1/2)\big(f_{M}(v_1;c_0,c_0) + f_{M}(v_1;-c_0,c_0)\big)f_{M}(v_2;0,c_0)f_{M}(v_3;0,c_0),
\end{align}
which will relax towards the Maxwellian distribution $f_{M}(\mathbf{v},\mathbf{0},\sqrt{(4/3)c_0^2})$. 
In figures (\ref{fig:gull})-(\ref{fig:tiger}) we show how the variance reduced estimator performs against the standard estimator, when estimating $\langle \left| v_1\right| \rangle$ using 100 particles, with and without the KDE stabilisation procedure. In both cases, the variance of the new estimator is smaller than the standard estimator, but the estimator with stabilisation from the KDE reduces the variance even further.

\subsection{Couette Flow}
\begin{figure}[h!]
  \centerline{\includegraphics[scale=0.5]{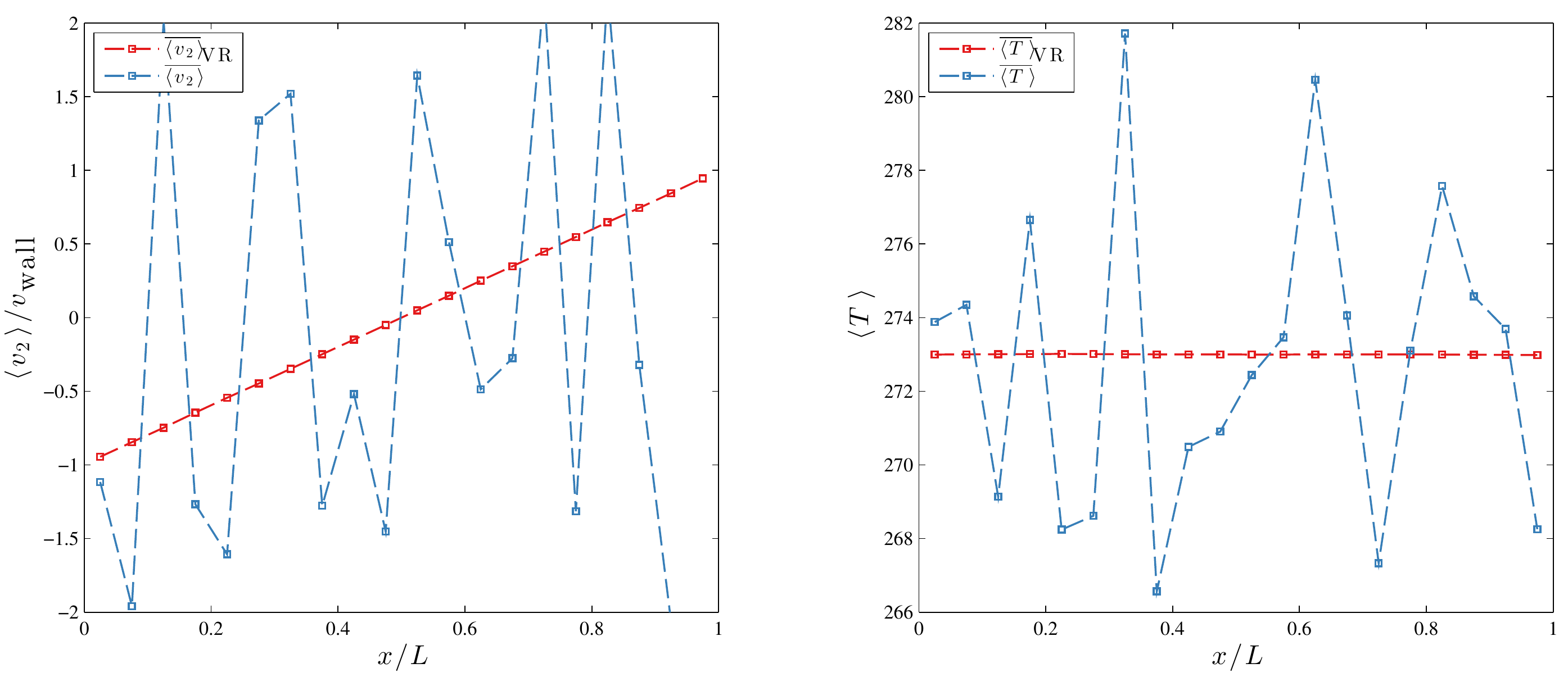}}
  \caption{Couette flow with wall velocity $v_{\textrm{wall}} = 0.01c_0$, $\textrm{Kn} = 0.05$, 20 cells and 100 particles per cell. }
  \label{fig: couette1}
\end{figure}

\begin{figure}[h!]
  \centerline{\includegraphics[scale=0.5]{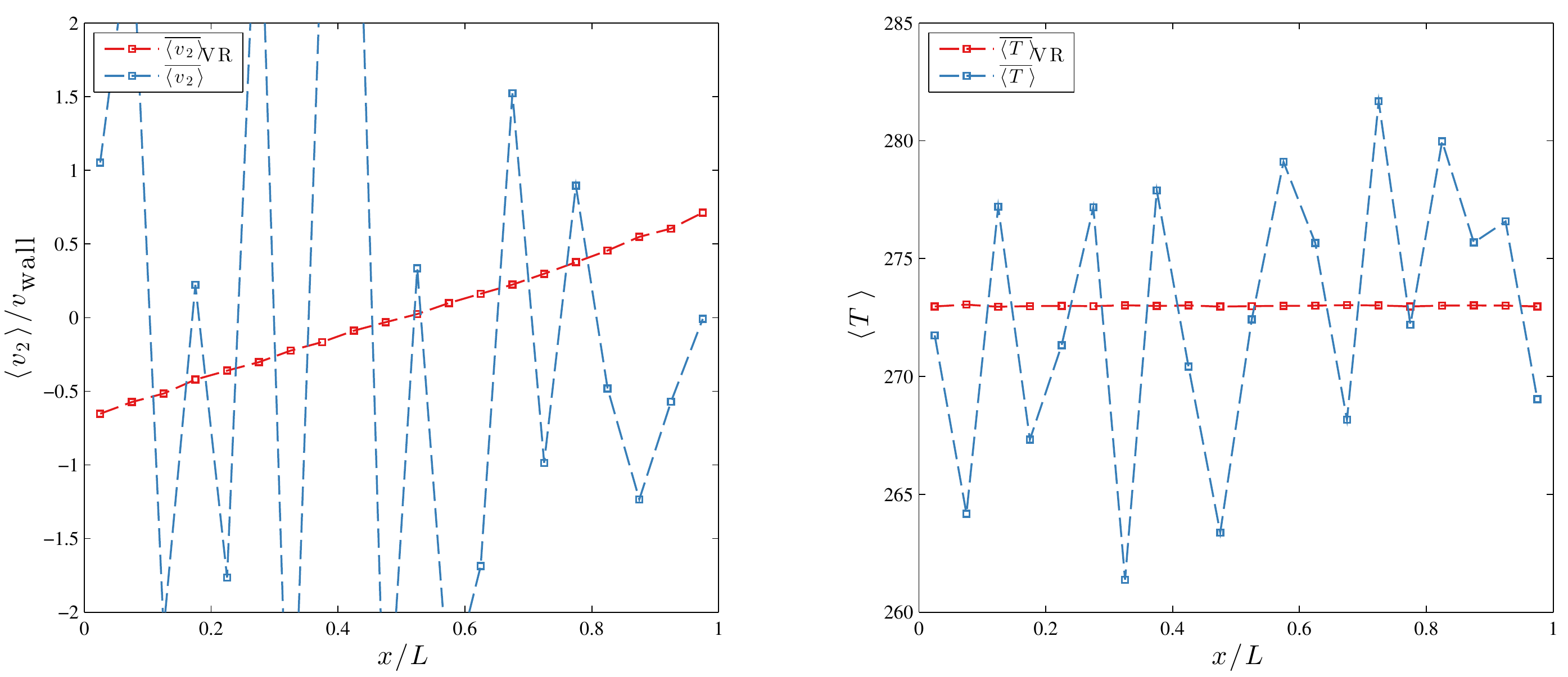}}
  \caption{Couette flow with wall velocity $v_{\textrm{wall}} = 0.01c_0$, $\textrm{Kn} = 0.5$, 20 cells and 100 particles per cell. }
  \label{fig: couette2}
\end{figure}

\begin{figure}[h!]
  \centerline{\includegraphics[scale=0.5]{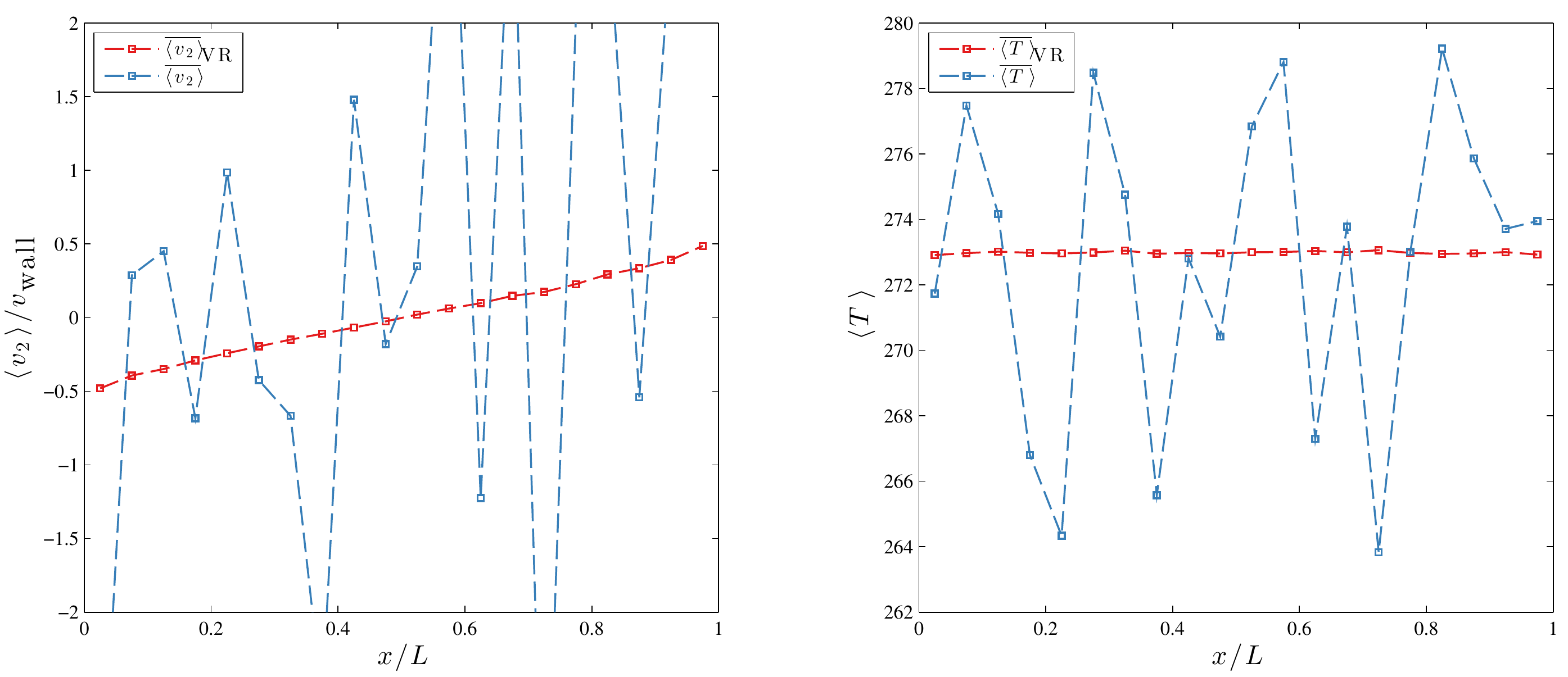}}
  \caption{Couette flow with wall velocity $v_{\textrm{wall}} = 0.01c_0$, $\textrm{Kn} = 1.0$, 20 cells and 100 particles per cell. }
  \label{fig: couette3}
\end{figure}

To test the particle weight variance reduction, we have applied the scheme to sample from a steady-state planar Couette flow, and compare to results obtained using a common random number scheme. A Couette flow is a flow where the fluid is bounded by two parallel walls moving in opposite directions within their planes, with velocity $\pm U_{\text{wall}}$. For Knudsen numbers $Kn=0.05,0.5,1.0$ respectively, Figures (\ref{fig: couette1}), (\ref{fig: couette2}), (\ref{fig: couette3}) show the variance reduced and standard Monte Carlo estimators of the steady-state flow velocity field parallel to the wall, $v_2(x_1)$, (left) as well as the temperature profile across the channel $T(x_1)$, for a Couette flow with wall velocity $v_{\textrm{wall}} = 0.01c_0$, $\textrm{Kn} = 0.5$, 20 cells and 100 particles per cell. All the results show a significant improvement of performance over the unweighted standard Monte Carlo estimator.

\begin{figure}[h!]
  \centerline{\includegraphics[scale=0.8]{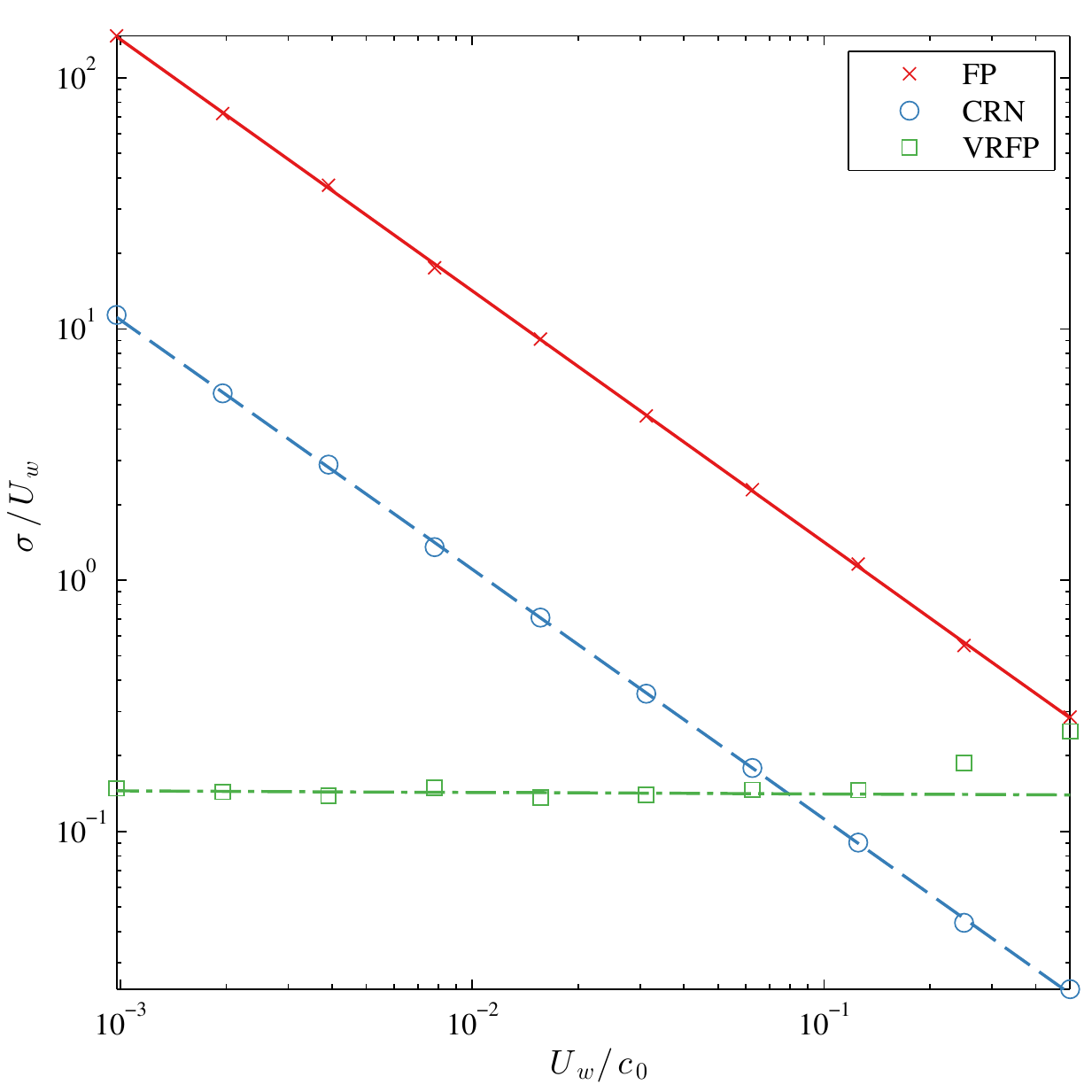}}
  \caption{Comparison of noise-to-signal ratio vs signal size, between standard Monte Carlo, CRN, and our importance sampling method (VRFP). }
  \label{fig:comparison}
\end{figure}

Next we compare the VRFP importance sampling scheme to the CRN correlated equilibrium scheme. Because we are interested in the noise of the estimate of the velocity profile across the channel, and the speed of the flow is small, we make the simplifying assumption that the steady-state density across the channel is constant. This allows us to choose the coefficients $\mathbf{A} = \mathbf{z}/\tau$ and $D= \sqrt{2RT_{wall}/\tau}$ so that the correlated equilibrium process is distributed according to the global Maxwellian,

\begin{align}
f_M(\mathbf{x},\mathbf{z},t) = \frac{\rho}{\left(2\pi RT_{wall}\right)^{3/2}}\exp\left(\frac{-z^2}{2RT_{wall}}\right)
\end{align} 
Figure \ref{fig:comparison}  compares the noise-to-signal ratio of the CRN scheme, VRFP scheme and standard Monte Carlo estimator against signal strength for the samples taken from steady state Couette flow estimator. The results show that for the CRN scheme there is a reduction in the noise-to-signal ratio by a factor of 10 over all tested levels of signal, corresponding to a speed up of over 100 times. However, it suffers the same scaling properties with signal size as the standard Monte Carlo estimator. In contrast, the importance weighted variance reduced estimator has a noise-to-signal ratio that is independent of the signal size as the signal size decreases. This results in an unbounded speed-up over the standard Monte-Carlo estimator as the signal size decreases to zero. Because of the independence of the signal strength on the noise-to-signal ratio, there is a signal strength where for larger signal strengths, the CRN scheme outperforms the particle weight scheme, and for Couette flow we estimate this to be at a Mach number greater than $0.1$.

\subsection{Lid-Driven Cavity}

To further demonstrate the effectiveness of the method, we apply it to a lid-driven cavity flow, where the fluid is bounded in two dimensions by a square box in the $(x,y)$ plane, with translational symmetry in the $z$ axis. Three of the bounding walls are stationary, and one of the bounding walls moves within its plane at constant velocity $U_{\text{wall}}$, giving rise to a circulatory flow within the cavity. 

\begin{figure}[h!]
        \centering
\begin{subfigure}[b]{0.495\textwidth}
\includegraphics[width=\textwidth]{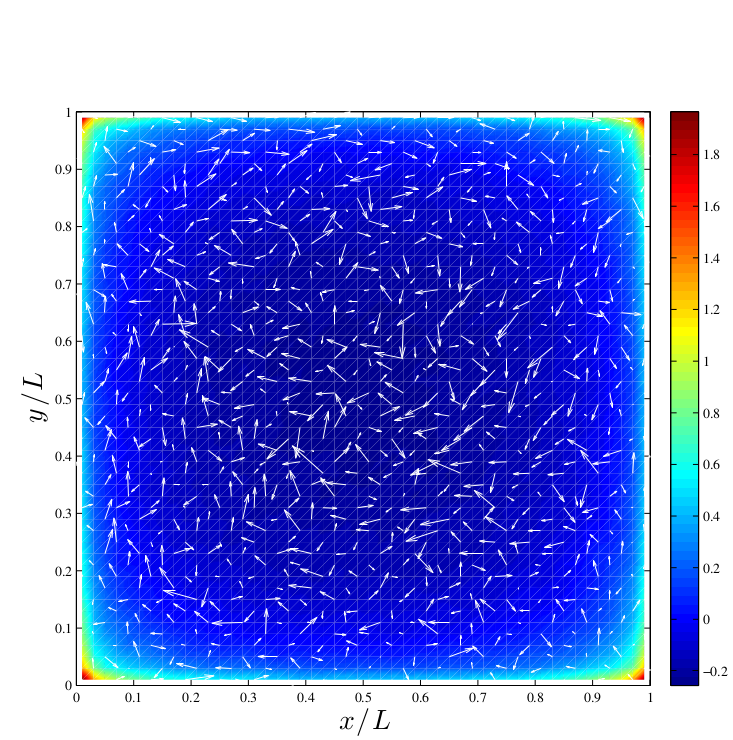}
 \caption{Standard Monte Carlo estimate}
  \label{fig:ldci}
  \end{subfigure}
          \begin{subfigure}[b]{0.495\textwidth}
\includegraphics[width=\textwidth]{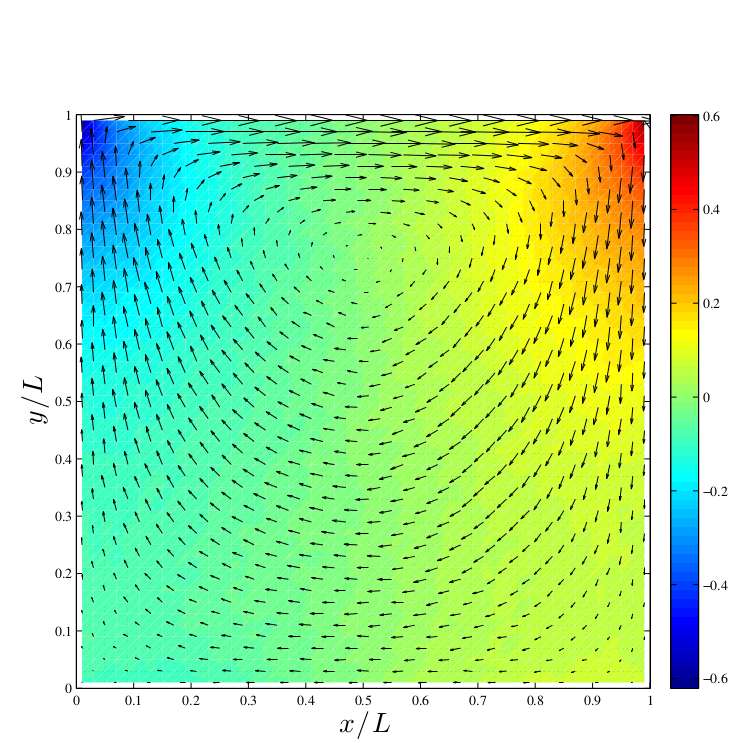}
  \caption{Variance reduced estimate}
  \label{fig:ldcii}
  \end{subfigure}
  \caption{Lid-driven cavity flow. Velocity field and non-dimensional temperature $(T/T_0 - 1)/\text{Ma}$. $\text{Kn}=1.0$, $U_{\text{wall}}=0.001c_0$, with and without importance sampling variance reduction. }
\end{figure}

\begin{figure}[h!]
  \centerline{\includegraphics[scale=0.8]{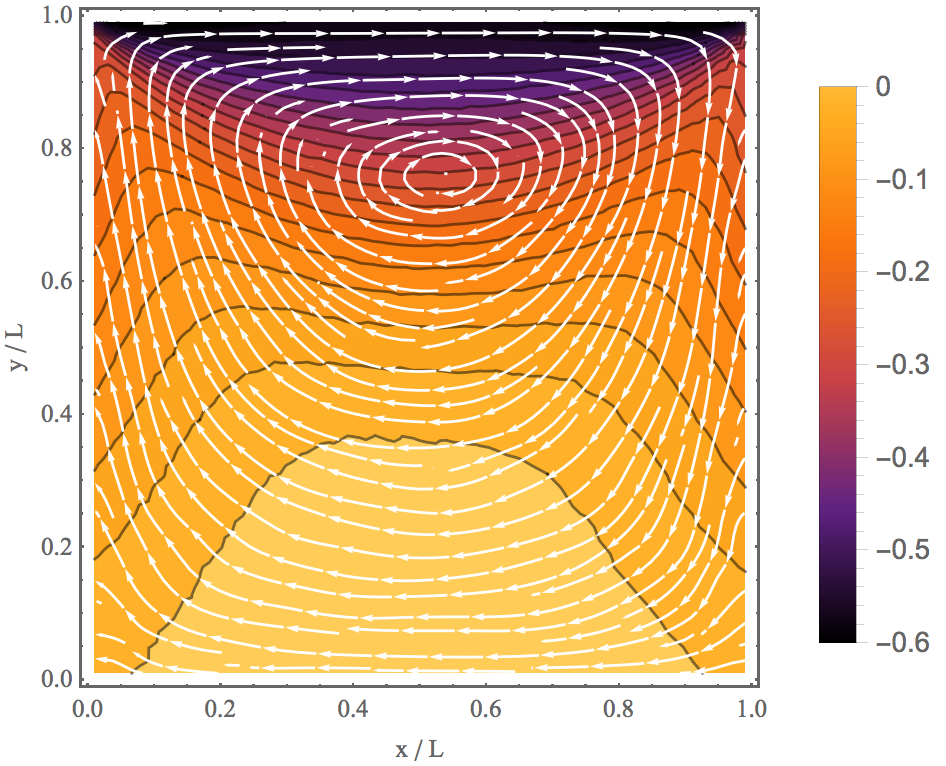}}
  \caption{Lid-driven cavity flow. Streamlines and non-dimensional sheer stress $\pi_{12}/(\rho_0 R T_0 \text{Ma})$ of the variance reduced estimate. }
  \label{fig:sheer}
\end{figure}

Figures (\ref{fig:ldci})-(\ref{fig:ldcii}) show the velocity and non-dimensional temperature field ($T/T_0 -1$) of the steady state flow, with a lid velocity of $U_{\text{wall}}=0.001c_0$ for the standard Monte Carlo and variance reduced sampling schemes. The results have been averaged over 5000 time-steps, and 10 independent ensembles on a $50\times50$ grid, with an average of 30 particles per cell. The standard Monte Carlo scheme is not able to pick up the signal, whereas we see clearly that the importance sampling scheme is able to recover the signal. In Figure \ref{fig:sheer} we compute the streamlines of the variance reduced flow, and the sheer stress $\pi_{12}$.

\begin{figure*}
\centering

\begin{subfigure}[t]{0.45\textwidth}
\centering
\includegraphics[width=\textwidth]{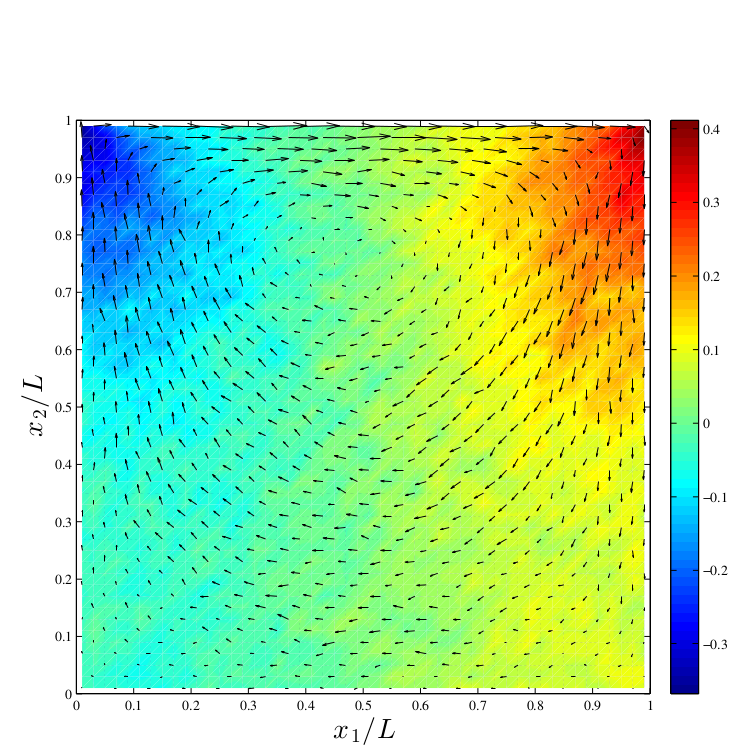}
\caption{ $U_{\text{wall}}=0.1c_0$}
\label{fig:mean and std of net14}
\end{subfigure}%
\hfill
\begin{subfigure}[t]{0.45\textwidth}
\centering
\includegraphics[width=\textwidth]{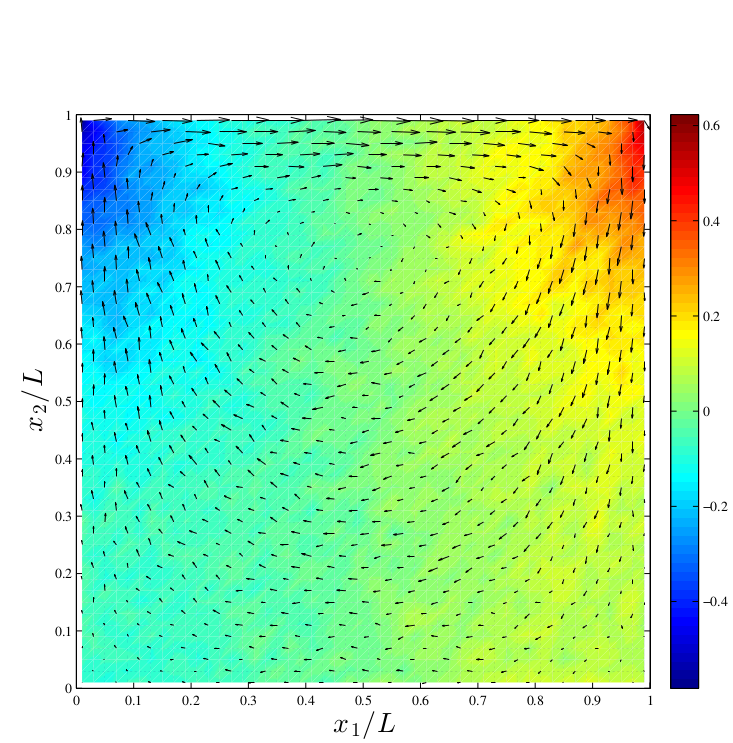}
\caption{$U_{\text{wall}}=0.01c_0$}
\label{fig:mean and std of net24}
\end{subfigure}

\bigskip 

\begin{subfigure}[t]{0.45\textwidth}
\centering
\includegraphics[width=\textwidth]{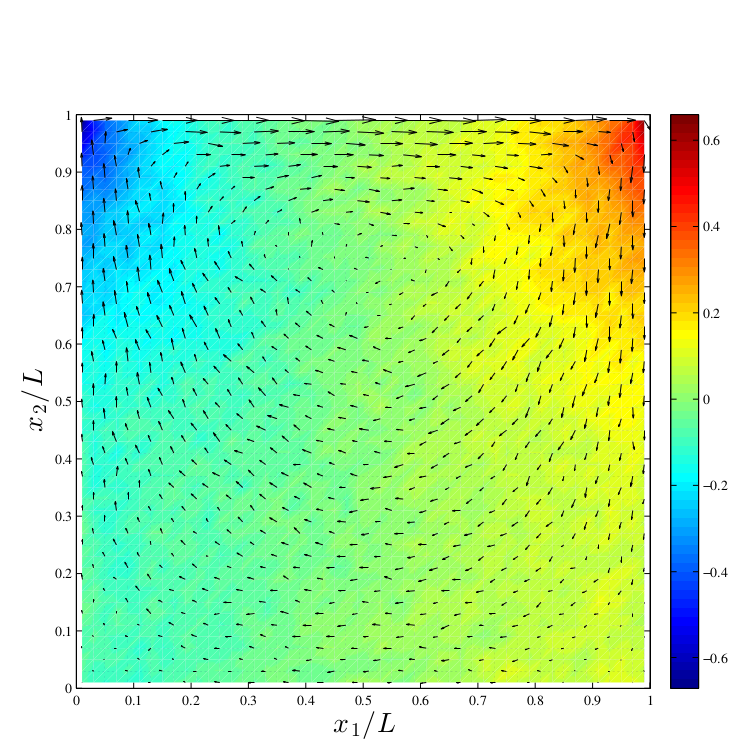}
\caption{$U_{\text{wall}}=0.001c_0$}
\label{fig:mean and std of net34}
\end{subfigure}%
\hfill
\begin{subfigure}[t]{0.45\textwidth}
\centering
\includegraphics[width=\textwidth]{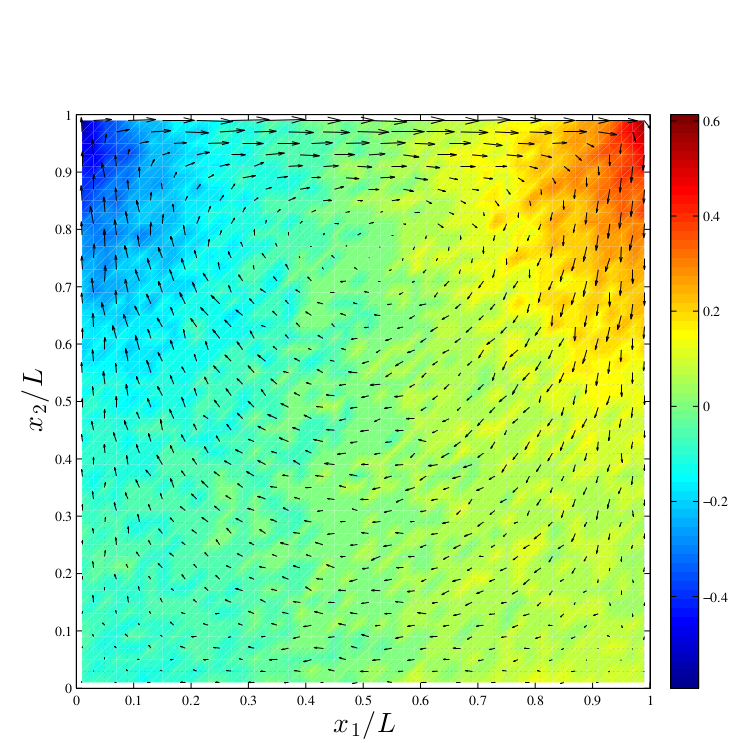}
\caption{$U_{\text{wall}}=0.0001c_0$}
\label{fig:mean and std of net44}
\end{subfigure}

\caption{Lid-driven cavity flows with different wall-speeds. $50\times50$ grid, $25$ particles per cell on average, $5000$ time steps to reach steady-state, thermodynamic fields averaged from $5000$ further time steps. The level of noise is independent to the wall-speed. }
\label{fig:kncomp}
\end{figure*}

Figure \ref{fig:kncomp} shows results from lid-driven cavity flows with lid speeds $0.1c_0$, $0.01c_0$, $0.001c_0$ and $0.0001c_0$. As was the case with the Couette flow, the level of noise in each calculation is independent of the lid-speed.

\section{Conclusion}

In this paper we have developed an importance sampling method for the Fokker-Planck rarefied gas model, that assigns weights to each stochastic particle allowing one to sample from an equilibrium distribution. We have demonstrated its effectiveness in reducing the variance of estimates of thermodynamics quantities for low Mach number flows over a range of Knudsen numbers. Significantly, the level of noise in the estimators becomes independent of the Mach number for low-speed flows. We believe it to be a versatile and robust method, and because it doesn't alter the basic algorithm of the particle solution scheme, it is able to be used in conjunction with other variance reduction schemes such as the CRN method.

\section{Aknowledgements}

We thank Dr. Hossein Gorji and coworkers for sending us their paper \citep{Gorji2015} prior to its finalized publication. This research is financially supported by EPSRC Programme Grant EP/I011927/1.


\bibliography{new_ref}






\end{document}